\documentclass{emulateapj}
\usepackage{color} % For use in the macro \query defined below

\usepackage{epsfig}
\usepackage{subfigure}
\usepackage{graphicx}
\usepackage{amsmath}
\usepackage{natbib}
\newcommand{\pa}{\partial}
\newcommand{\mb}{\boldsymbol}

\shorttitle{Accretion in PPDs with Tiny Grains}
\shortauthors{X.-N. Bai}

\begin{document}

%% LaTeX will automatically break titles if they run longer than
%% one line. However, you may use \\ to force a line break if
%% you desire.

\title{The Role of Tiny Grains on the Accretion Process in Protoplanetary Disks}

%% Use \author, \affil, and the \and command to format
%% author and affiliation information.
%% Note that \email has replaced the old \authoremail command
%% from AASTeX v4.0. You can use \email to mark an email address
%% anywhere in the paper, not just in the front matter.
%% As in the title, use \\ to force line breaks.

\author{Xue-Ning Bai}
\affil{Department of Astrophysical Sciences, Princeton University,
Princeton, NJ, 08544} \email{xbai@astro.princeton.edu}

%% Notice that each of these authors has alternate affiliations, which
%% are identified by the \altaffilmark after each name.  Specify alternate
%% affiliation information with \altaffiltext, with one command per each
%% affiliation.

%\altaffiltext{2}{Society of Fellows, Harvard University.}

%% Mark off your abstract in the ``abstract'' environment. In the manuscript
%% style, abstract will output a Received/Accepted line after the
%% title and affiliation information. No date will appear since the author
%% does not have this information. The dates will be filled in by the
%% editorial office after submission.

\begin{abstract}
Tiny grains such as polycyclic aromatic hydrocarbons (PAHs) have been thought to
dramatically reduce the coupling between gas and magnetic fields in weakly ionized gas
such as in protoplanetary disks (PPDs) because they provide tremendous surface area to
recombine free electrons. The presence of tiny grains in PPDs thus raises the question of
whether the magnetorotational instability (MRI) is able to drive rapid accretion to be consistent
with observations.
Charged tiny grains have similar conduction properties as ions, whose presence leads to
qualitatively new behaviors in the conductivity tensor, characterized by $\bar{n}/n_e>1$, where
$n_e$ and $\bar{n}$ denote the number densities of free electrons and all other charged
species respectively. In particular, Ohmic conductivity becomes dominated by charged grains
rather than electrons when $\bar{n}/n_e$ exceeds about $10^3$, and Hall and ambipolar
diffusion (AD) coefficients are reduced by a factor of $(\bar{n}/n_e)^2$ in the AD dominated
regime relative to that in the Ohmic regime.
Applying the methodology of Bai (2011), we find that in PPDs, when PAHs are sufficiently
abundant ($\gtrsim10^{-9}$ per H$_2$ molecule), there exists a transition radius
$r_{\rm trans}$ of about $10-20$ AU, beyond which the MRI active layer extends to the disk
midplane. At $r<r_{\rm trans}$, the optimistically predicted MRI-driven accretion rate $\dot{M}$
is one to two orders of magnitude smaller than that in the grain-free case, which is too small
compared with the observed rates, but is in general no smaller than the predicted $\dot{M}$ with
solar-abundance $0.1\mu$m grains. At $r>r_{\rm trans}$, we find that
remarkably, the predicted $\dot{M}$ exceeds the grain-free case due to a net reduction of AD
by charged tiny grains, and reaches a few times $10^{-8}M_{\bigodot}$ yr$^{-1}$. This is
sufficient to account for the observed $\dot{M}$ in transitional disks. Larger grains
($\gtrsim0.1\mu$m) are too massive to reach such high abundance as tiny grains and to
facilitate the accretion process.
\end{abstract}

%% Keywords should appear after the \end{abstract} command. The uncommented
%% example has been keyed in ApJ style. See the instructions to authors
%% for the journal to which you are submitting your paper to determine
%% what keyword punctuation is appropriate.


\section{Introduction}\label{sec:intro}

Weakly ionized gas is not perfectly conducting. In the presence of magnetic field (${\mb B}$),
the Ohm's law is anisotropic, giving rise to Ohmic resistivity, Hall effect, and ambipolar
diffusion (AD) as the three non-ideal magnetohydrodynamics (MHD) effects. More formally,
assuming the inertia in the ionized species is negligible, the dynamical equations for the
neutrals are the same as ideal MHD except for the induction equation, which reads
\begin{equation}
\frac{\pa{\mb B}}{\pa t}=\nabla\times({\mb v}\times{\mb B})
-\frac{4\pi}{c}\nabla\times[\eta_O{\mb J}+\eta_H({\mb J}\times{\hat{\mb B}})
+\eta_A{\mb J}_\perp]\ ,\label{eq:induction}
\end{equation}
where ${\mb v}$ is the velocity of the neutrals, ${\mb J}=(c/4\pi)\nabla\times{\mb B}$ is the
current density, $\hat{\mb B}$ denotes unit vector along ${\mb B}$, subscript ``$_\perp$"
denotes the vector component that is perpendicular to ${\mb B}$, $\eta_O, \eta_H$ and
$\eta_A$ are the Ohmic, Hall and the ambipolar diffusivities, which depend on the number
density of the charged species, and the latter two also depend on $B$. When ions and
electrons are the only charged species in the gas (which can be relaxed to when the electrons
are the main negative charge carriers), a simple expression for the diffusion coefficients reads
\citep{SalmeronWardle03}:
\begin{equation}
\eta_O=\eta_e\ ,\qquad\eta_H=\frac{cB}{4\pi en_e}\ ,\qquad
\eta_A=\frac{B^2}{4\pi\gamma_i\rho\rho_i}\ ,\label{eq:diff0}
\end{equation}
where $\eta_e$ is the resistivity due to the electrons, $n_e$ is the electron number density,
$\rho$ and $\rho_i$ are the mass density of the neutrals and the ions, $\gamma_i$ is defined
after equation (\ref{eq:betaj}). In general, Ohmic effect dominates in dense regions with weak
magnetic field, AD dominates in tenuous regions with strong magnetic field, while the Hall
regime lies in between. 

Grains, which typically contribute to about $\lesssim1\%$ of mass in the gas, can possess both
positive and negative charges. Grains are well known for significantly enhancing the electron
recombination rate \citep{DraineSutin87}, which reduces the ionization fraction ($n_e/n_H$,
where $n_H$ is the number density of hydrogen atom), and leads to greatly increased Ohmic
resistivity. However, in sufficiently weakly ionized gas, the abundance of tiny grains (e.g., the
polycyclic aromatic hydrocarbon, PAH) may well exceed the electron abundance without
contributing much to the mass budget. In this paper, we generalize equation (\ref{eq:diff0}) to
include positive and negatively charged tiny grains and show that for sufficiently weak ionization,
tiny grains can dramatically reduce $\eta_H$ and $\eta_A$ as the main charge carrier switches
to the grains instead of ions and electrons.

Our result is directly applicable to protoplanetary disks (PPDs), where the ionization fraction
resulting from major ionization sources such as X-rays and cosmic rays is generally orders of
magnitude below unity, and it decreases from surface to the midplane because of the
attenuation of ionizing particles \citep{Gammie96}. Non-ideal MHD effects are closely relevant
to the magnetorotational instability (MRI, \citealp{BH91}), whose linear dispersion relation as
well as the non-linear saturation properties are strongly affected \citep{Wardle99,Balbus09},
and whether the MRI is responsible for driving rapid accretion with
$\dot{M}\sim10^{-8\pm1}M_{\bigodot}$ yr$^{-1}$ in PPDs \citep{Hartmann_etal98} has been a
long-standing problem. Most studies have focused on the role of the Ohmic resistivity (e.g.,
\citealp{Turner_etal07,BaiGoodman09}),
while it was not until recently have AD been taken into account to estimate the effectiveness of
the MRI \citep{CMC07,PerezBeckerChiang11}. Combining the most recent results from
numerical simulations of the MRI with AD \citep{BaiStone11a}, \citet{Bai11a} showed that while
the MRI can always operate in PPDs, AD can be a main limiting factor for the MRI-driven
accretion (besides Ohmic resistivity). The reduction of AD coefficient by tiny grains (e.g., PAHs),
while counterintuitive, helps enhance the accretion rate, as we will demonstrate in this paper.

Although observational data show strong evidence of grain growth to micron size or larger in
PPDs (e.g., \citealp{DAlessio_etal01,vanBoekel_etal03}), PAH emission has also been detected
in majority of Herbig Ae/Be stars \citep{AckeAncker04}, as well as a small fraction of T-Tauri stars
\citep{Geers_etal06,Oliveira_etal10}. As argued in \citet{PerezBeckerChiang11}, PAHs may be
equally abundant in T-Tauri disks but they fluoresce less luminously due to fainter ultraviolet
radiation field of their host stars. The existence of PAHs also suggests a continuous size
distribution of grains to the smallest end of a few ${\rm \AA}$ as a result of grain coagulation and
fragmentation. Throughout this paper, we use the phrases ``tiny grain" and PAH interchangeably,
which refer to grain with size $a\lesssim0.01\mu$m. Tiny grains may dominate larger grains in
abundance while contribute a negligible fraction of the total grain mass. Taking the grain density to
be 3 g cm$^{-3}$ and the gas mean molecular weight $\mu_n$ to be $2.34$ atomic mass (as
appropriate for PPDs), the relation between grain mass fraction ($f$) and grain abundance per
H$_2$ molecule ($x$) reads
\begin{equation}
\frac{f}{0.01}\approx0.32\bigg(\frac{x}{10^{-9}}\bigg)\bigg(\frac{a}{0.01\mu{\rm m}}\bigg)^{3}\ ,
\end{equation}
where $a$ is grain size. The abundance of grains with $a\lesssim0.01\mu$m may easily
reach large abundance of $10^{-9}$ or higher, while the abundance of $a\gtrsim0.1\mu$m
grains would be at most about $10^{-12}$. The significance of this abundance cut ($x=10^{-9}$)
will be addressed in this paper.

In Section \ref{sec:grain} we describe a generalized model for non-ideal MHD diffusion
coefficients with the inclusion of charged grains. The model is applied to interpret the the
results in Section \ref{sec:ppd}, where we study the role of tiny grains in PPDs following the
methodology of \citet{Bai11a}. Summary and discussion follow in Section \ref{sec:conclusion}.

\section[]{Non-ideal MHD Effects with Grains}\label{sec:grain}

The Ohm's Law derives from the motion of charged particles, which is characterized
by the Hall parameter, the ratio between the gyrofrequency and the momentum
exchange rate \citep{Wardle07}. For species $j$ with mass $m_j$ and charge $Z_je$, the
Hall parameter reads
\begin{equation}
\beta_j\equiv\frac{|Z_j|eB}{m_jc}\frac{1}{\gamma_j\rho}\ .\label{eq:betaj}
\end{equation}
where $\gamma_j\equiv\langle\sigma v\rangle_j/(\mu_n+m_j)$ with $\langle\sigma v\rangle_j$
being the rate coefficient for momentum transfer between charged species $j$ with the
neutrals and $\mu_n$ is the mean molecular weight of the neutrals. Charged species $j$ is
strongly coupled to the neutrals if $|\beta_j|\ll1$, and is strongly tied to magnetic fields when
$|\beta_j|\gg1$.

In weakly ionized gas with a number of different charged species, the general expressions
for the Ohmic, Hall and ambipolar diffusion coefficients are given by \citep{Wardle07,Bai11a}
\begin{equation}
\begin{split}
\eta_O&=\frac{c^2}{4\pi\sigma_O}\ ,\\
\eta_H&=\frac{c^2}{4\pi\sigma_\perp}\frac{\sigma_H}{\sigma_\perp}\ ,\\
\eta_A&=\frac{c^2}{4\pi\sigma_\perp}\frac{\sigma_P}{\sigma_\perp}-\eta_O\ ,\\
\end{split}\label{eq:eta_general}
\end{equation}
where $\sigma_\perp\equiv\sqrt{\sigma_H^2+\sigma_P^2}$ and the Ohmic, Hall and
Pedersen conductivities are
\begin{equation}
\begin{split}
\sigma_O&=\frac{ec}{B}\sum_jn_j|Z_j|\beta_j\ ,\\
\sigma_H&=\frac{ec}{B}\sum_j\frac{n_jZ_j}{1+\beta_j^2}\ ,\\
\sigma_P&=\frac{ec}{B}\sum_j\frac{n_j|Z_j|\beta_j}{1+\beta_j^2}
\end{split}\label{eq:sigma_general}
\end{equation}
respectively, where the summation goes over all charged species. Note that the Hall conductivity
depends on the sign of $Z_j$, while $\sigma_O$ and $\sigma_P$ depend only on $|Z_j|$.

The simple expressions for the magnetic diffusion coefficients (\ref{eq:diff0}) can be obtained by
assuming electrons and ions are the only two charged species. Below we generalize it to include
tiny grains. Besides the reason mentioned in the introduction, we consider tiny grains because
they are unlikely to possess multiple charges due to higher potential barrier
\citep{PerezBeckerChiang11}, which simplifies the algebra considerably. In addition, the mean
free path for tiny grains is sufficiently small so that they can be treated as fluid.

The momentum transfer rate coefficients for electrons, ions and grains can be found in
equations (14) - (16) of \citet{Bai11a}. Converting to the Hall parameter, we have
\begin{equation}
\begin{split}
\beta_i&\approx3.3\times10^{-3}\frac{B_G}{n_{15}}\ ,\\
\beta_e&\approx2.1\frac{B_G}{n_{15}}\max\bigg[1,\bigg(\frac{T}{100{\rm K}}\bigg)^{1/2}\bigg]
\gg\beta_i\ ,\\
\end{split}
\end{equation}
for ions and electrons respectively, where $B_G$ is the magnetic field strength measured in
Gauss, $n_{15}=n_H/10^{15}$ cm$^{-3}$, and the $\mu_n$ is taken to be $2.34$ atomic mass
appropriate for PPDs. For singly charged small grains, the resulting $\beta$ is very close to
$\beta_i$, because the reduced gyro-frequency due to larger mass is compensated by the
reduced momentum transfer rate due to larger inertia. The electron Hall parameter $\beta_e$ is
much larger than $\beta_i$ because electrons are the most mobile species. Therefore, in terms
of conductivity, we only have two groups of charges: electrons with Hall parameter $\beta_e$,
and ions / charged grains with Hall parameter $\beta_i$.

Let $\bar{n}=n_i+n_{\rm gr}^++n_{\rm gr}^-$ be the total number density of the second group. 
Since grains can carry negative charge, one has $\bar{n}\geq n_e$. Generalization from
$\bar{n}=n_e$ (grain-free case, which leads to equation (\ref{eq:diff0})) to $\bar{n}>n_e$
involves more complicated algebra, as we describe below. For brevity, we ignore the pre-factor
$en_ec/B$ in conductivities and $cB/4\pi en_e$ in diffusivities, which make them
dimensionless.

The three components of the conductivity tensor read
\begin{equation}
\begin{split}
\sigma_O&=\beta_e+\frac{\bar{n}}{n_e}\beta_i=(\beta_e+\beta_i)(1+\theta)\ ,\\
\sigma_H&=\frac{1}{1+\beta_e^2}-\frac{1}{1+\beta_i^2}
=\frac{(\beta_e+\beta_i)(\beta_e-\beta_i)}{(1+\beta_e^2)(1+\beta_i^2)}\ ,\\
\sigma_P&=\frac{\beta_e}{1+\beta_e^2}+\frac{(\bar{n}/n_e)\beta_i}{1+\beta_i^2}
=\frac{(\beta_e+\beta_i)}{(1+\beta_i^2)}\bigg[\frac{(1+\beta_e\beta_i)}{(1+\beta_e^2)}
+\theta\bigg]\ ,
\end{split}\label{eq:sigma_general}
\end{equation}
where we have defined
\begin{equation}
\theta\equiv\frac{\bar{n}-n_e}{n_e}\frac{\beta_i}{\beta_e+\beta_i}
\approx\frac{n_{\rm gr}^{\pm}\beta_i}{n_e\beta_e}\ .
\end{equation}
The parameter $\theta$ is independent of magnetic field, and measure the ratio of grain
conductivity to electron conductivity. Note that $\theta=0$ in the grain-free case, and the
Hall conductivity is independent of $\theta$. The perpendicular conductivity is
\begin{equation}
\sigma_\perp^2=\frac{(\beta_e+\beta_i)^2}{(1+\beta_e^2)(1+\beta_i^2)}\frac{1}{f(\theta)}\ ,
\end{equation}
where
\begin{equation}
f(\theta)\equiv\frac{(1+\beta_i^2)}{[(1+\theta)^2+(\beta_i+\theta\beta_e)^2]}\ .
\end{equation}
Note that $f(\theta)=1$ in the grain-free case ($\theta=0$). It is straightforward to obtain
the Ohmic, Hall and ambipolar diffusion coefficients, after some algebra
\begin{equation}
\eta_O=\frac{1}{(\beta_e+\beta_i)(1+\theta)}\approx\frac{1}{\beta_e(1+\theta)}\ ,
\end{equation}
\begin{equation}
\eta_H=\frac{\beta_e-\beta_i}{\beta_e+\beta_i}f(\theta)\approx f(\theta)\ ,
\end{equation}
\begin{equation}
\begin{split}
\eta_A&=\frac{[(1+\theta)+\beta_e(\beta_i+\theta\beta_e)]}{(\beta_e+\beta_i)}f(\theta)
-\eta_O\\
&\approx\bigg[(1+\theta)(\beta_i+\theta\beta_e)-\frac{\beta_e\theta^2}{1+\beta_i^2}\bigg]
\frac{f(\theta)}{1+\theta}\ ,
\end{split}
\end{equation}
where the approximate formulae are obtained by noting that $\beta_e\gg\beta_i$, with error on the
order of $\beta_i/\beta_e\sim10^{-3}$.

We see that the grains affect the magnetic diffusion coefficients mainly via two factors: $(1+\theta)$
and $(\beta_i+\theta\beta_e)/\beta_i\approx\bar{n}/n_e$, both of which are independent of magnetic
field strength. The former describes the ratio of total conductivity to electron conductivity, while the
latter describes the number density ratio of ions / grains to electrons, and is much more substantial
than the factor (1+$\theta$) since $\beta_e\gg\beta_i$.

For $\theta\ll\beta_i/\beta_e\sim10^{-3}$, we generally have $\bar{n}\approx n_i\approx n_e$,
$f(\theta)\approx1$, and the classical results (\ref{eq:diff0}) are recovered. In particular, Ohmic
resistivity dominates when $\beta_e<1$, where both ions and electrons are coupled to the neutrals;
Hall regime corresponds to $\beta_i<1<\beta_e$, where electrons are coupled to the magnetic field
while the ions are not; AD regime corresponds to $\beta_i>1$ where both electrons and ions are
coupled to the magnetic field. This is the classical interpretation of the non-ideal MHD effects.

Qualitatively new behaviors appear when $\theta\gtrsim\beta_i/\beta_e$, where grains play a
dominant role in the abundance of charged particles ($n_{\rm gr}^{\pm}\gtrsim n_e$). In the next
two paragraphs, we discuss the asymptotic behaviors of the diffusion coefficients, where
comparison is made with the grain-free diffusion coefficients (\ref{eq:diff0}) at fixed $n_e$. In
reality, tiny grains strongly reduces $n_e$ relative to the grain-free case, which will be discussed
at the end of this section.

The Ohmic resistivity is least affected by grains, which is simply reduced by a factor of
$(1+\theta)$ relative to the electron resistivity regardless of magnetic field strength. Therefore,
grain conductivity becomes important when $\theta\gtrsim1$ or $\bar{n}\gtrsim10^3n_e$. For
Hall and AD coefficients, we consider two separate limits. When $\beta_i\ll\beta_e\ll1$ (Ohmic
regime), we have
\begin{equation}
\begin{split}
\eta_H&\approx\frac{cB}{4\pi en_e}\frac{1}{(1+\theta)^2}\ ,\\
\eta_A&\approx\frac{cB\beta_i}{4\pi en_e}\frac{\bar{n}}{n_e(1+\theta)^3}\ ,
\end{split}
\end{equation}
where we have factored out the results into the grain-free expression (left) multiplied by a correction
factor (right). We see that the Hall diffusivity is only moderately affected by grains, being reduced by
a factor of $(1+\theta)^2$ at fixed $n_e$. The AD, on the other hand, is enhanced by a factor of
$\bar{n}/n_e$, in addition to a moderate reduction by $(1+\theta)^{3}$. Nevertheless, changes in
Hall and AD coefficients do not play a significant role here since Ohmic resistivity is still the
dominant effect.

In the opposite limit $1\ll\beta_i\ll\beta_e$ (AD regime), we have
\begin{equation}
\begin{split}\label{eq:eta_AD}
\eta_H&\approx\frac{cB}{4\pi en_e}\bigg(\frac{n_e}{\bar{n}}\bigg)^2\ ,\\
\eta_A&\approx\frac{cB\beta_i}{4\pi en_e}\frac{n_e}{\bar{n}}\ .
\end{split}
\end{equation}
We see that in this limit and at fixed $n_e$, the Hall effect is reduced by a factor of
$(\bar{n}/n_e)^2$, while AD is reduced by a factor of $(\bar{n}/n_e)$. The reduction of AD is
easily understood. Without grains, the neutral gas is coupled to the magnetic field through
ion-neutral collisions, hence $\eta_A\propto1/n_i=1/n_e$. Electrons play a negligible role
because of its small inertia. Charged grains (no matter positive or negative) play exactly the
same role as ions, hence we have $\eta_A\propto1/\bar{n}$ in the presence of grains.
We also note that the Hall effect vanishes if positive and negative charge carriers have the
same mass, which is consistent with our result as $n_e/\bar{n}\rightarrow0$.

\begin{figure}
    \centering
    \includegraphics[width=90mm]{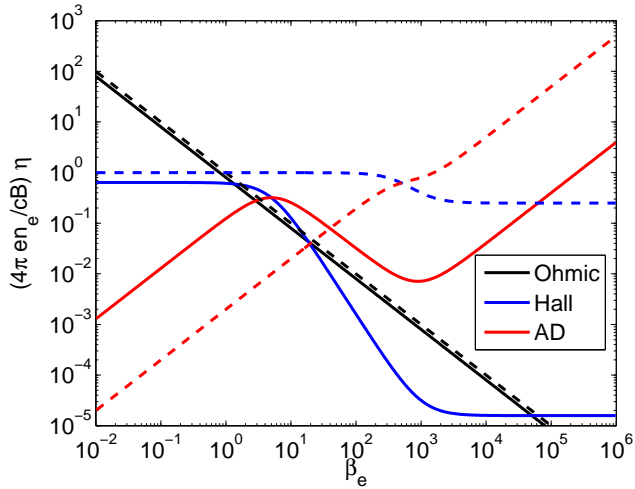}
  \caption{Dimensionless Ohmic (black), Hall (blue) and ambipolar (red) diffusion coefficients as
  a function of the electron Hall parameter $\beta_e=1000\beta_i$. We have considered two different
  values of the charged grain abundance parameter: $\theta=10^{-3}$ ($\bar{n}=2n_e$, dashed) and
  $\theta=0.25$ ($\bar{n}=251n_e$, solid).}\label{fig:model}
\end{figure}

In Figure \ref{fig:model} we show two sample calculations of the magnetic diffusion coefficients
for $\theta=10^{-3}$ (dashed) and $\theta=0.25$ (solid) respectively, which corresponds to
situations where charged grain abundance is comparable to, and greatly exceed the electron
abundance. The two asymptotic regimes derived above are clearly seen, with a transition region
in between. The $\theta=10^{-3}$ results are close to the grain-free case, where all three curves
are close to straight lines except weak transitions as $\beta_i$ passes $1$. For relatively large
$\theta$, the transition region extends from $\beta_e\approx1$ to $\beta_i\approx1$. Because of
the suppression of the Hall diffusion beyond $\beta_e\approx1$ and enhancement of AD before
$\beta_e\approx1$, AD becomes the dominant effect even when $\beta_i\ll1$ ($\beta_e\ll1000$),
and the Hall regime gradually diminishes as $\theta$ increases. These results nicely explain the
observed magnetic diffusivity pattern in Figure 2 of \citet{Bai11a}, as aided with the chemical
abundances shown in Figure 1 of the same paper.

Now let us take into account the chemistry in the gas. Assuming fixed gas density, temperature, and
ionization rate, the equilibrium electron number density in the presence of small grains $n_{e1}$ is
much smaller than that in the grain-free case $n_{e0}$ \citep{BaiGoodman09,PerezBeckerChiang11}.
Therefore, Hall and AD coefficients are made larger by a factor of $n_{e0}/n_{e1}$. However, this
factor is compensated by another reduction factor of $(n_{e1}/\bar{n})^2$ for the Hall effect and of
$n_{e1}/\bar{n}$ for AD in the AD dominated regime. Here we focus on AD. In order to have a net
reduction of AD coefficient when $\beta_i\gg1$ (AD regime), it is required that
\begin{equation}
\bar{n}>n_{e0}\ .\label{eq:criterion}
\end{equation}
That is, the charged grain abundance has to exceed the grain-free electron abundance. We will see
that this condition can be met in PPDs. Similarly, for Ohmic resistivity to be reduced relative to the
grain-free case, one requires $\bar{n}\gtrsim10^3n_{e0}$. This condition, however, is very unrealistic
given the disk chemistry (see Section \ref{ssec:chem}), thus Ohmic resistivity always increases in
the presence of tiny grains. Nevertheless, if condition (\ref{eq:criterion}) is met (which is generally
the case when PAHs are sufficiently abundant, see Section \ref{ssec:chem}), Ohmic resistivity is
enhanced by at most a factor of about $10^3$ relative to the grain-free case.

\section[]{Application to the Magnetorotational Instability in PPDs}\label{sec:ppd}

To illustrate the significance of the theoretical results in the previous section, we study the role of
tiny grains in PPDs. We refer to Section 3 of \citet{Bai11a} for all details about the methodology
and calculation procedures, which we follow exactly except that here we consider $a=1$ nm sized
grains (as a proxy for PAHs) instead of larger grains\footnote{We use binding energy of $D=3$ eV in
the calculation of the electron-grain sticking coefficient rather than 1 eV used in \citet{Bai11a}.}. The
PAH abundance in T-Tauri stars estimated by \citet{Geers_etal06} is about 
$x_{\rm PAH}\approx10^{-8}$-$10^{-7}$ per H$_2$ molecule, while \citet{PerezBeckerChiang11}
argued for smaller abundance due to dust settling. Moreover, the distribution of PAHs in PPDs
has been found to be spatially variable \citep{Geers_etal07}. For these reasons, we adopt
$x_{\rm PAH}=10^{-8}$ as fiducial, while we also
consider $x_{\rm PAH}$ down to $10^{-11}$ and up to $10^{-7}$. We fix the X-ray
luminosity to be $10^{30}$ erg s$^{-1}$, X-ray temperature to be 5 keV, and include cosmic-ray
ionization of $10^{-17}$ s$^{-1}$ with penetration depth of $96$ g cm$^{-2}$. We use the
minimum-mass solar nebular (MMSN, \citealp{Hayashi81}) as the fiducial disk model, while we also
consider the much more massive disk model by \citet{Desch07}. A complex chemical network of
more than $4000$ reactions are evolved at different heights and radii of the disk for 1 Myr to
(quasi-) chemical equilibrium, from which abundance of charged species are extracted to evaluate
the non-ideal MHD diffusion coefficients. We will mainly focus on the outer disk ($\gtrsim10$ AU)
where AD rather than Ohmic resistivity plays the dominant role the gas dynamics and the effect
discussed in Section \ref{sec:grain} is most relevant.

\begin{figure}
    \centering
    \includegraphics[width=90mm]{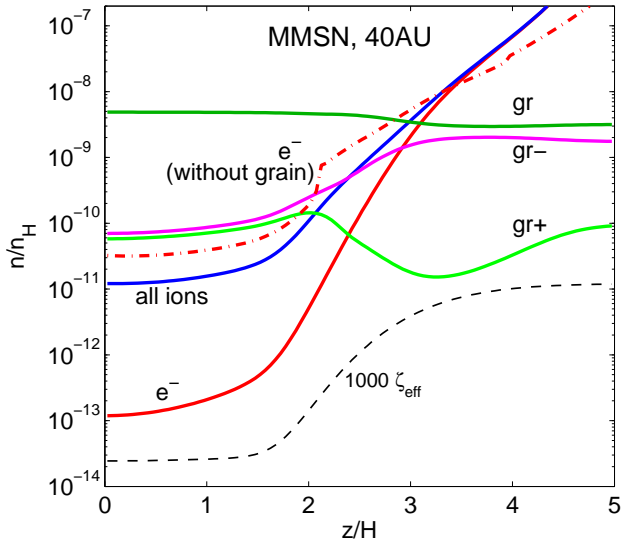}
  \caption{Abundance of charged species and grains (PAHs) as a function of disk height in our
  fiducial model calculation (MMSN disk with $x_{\rm PAH}=10^{-8}$) at 40 AU, plotted as solid lines
  with labels. For comparison, red dash-dotted line shows the electron (and ion) abundance in the
  grain-free calculation. The ionization rate is plotted (black dashed) for reference.}\label{fig:nsp}
\end{figure}

\subsection[]{Abundance of Charged Species}\label{ssec:chem}

In Figure \ref{fig:nsp} we show the electron, ion and grain abundances at $40$ AU in the MMSN
model as a function of disk height ($z$), normalized to the disk scale height ($H=c_s/\Omega$ where
$c_s$ is the sound speed, $\Omega$ is the disk angular frequency). We see that electrons and ions
are the dominant charged species (i.e., $\bar{n}\approx n_e$) above $\sim3H$, while below it
the abundances of ions and charged grains exceed electron abundance, and $\bar{n}/n_e$
reaches about $1000$ at the disk midplane. The large value of $\bar{n}/n_e$ implies strong
suppresion of AD and the Hall effect according to equation (\ref{eq:eta_AD}). To see its significance,
we also show the electron (thus ion) abundance in the grain-free calculation as the red dash-dotted
line in the Figure. Clearly, we see that $n_{e0}<\bar{n}$ for $z\lesssim2H$. According to equation
(\ref{eq:criterion}), this means that when tiny grains are present, the AD coefficient is even smaller
than that in the grain-free case! Seemingly counterintuitive, one can qualitatively understand this
result as follows.

\begin{figure*}
    \centering
    \includegraphics[width=160mm]{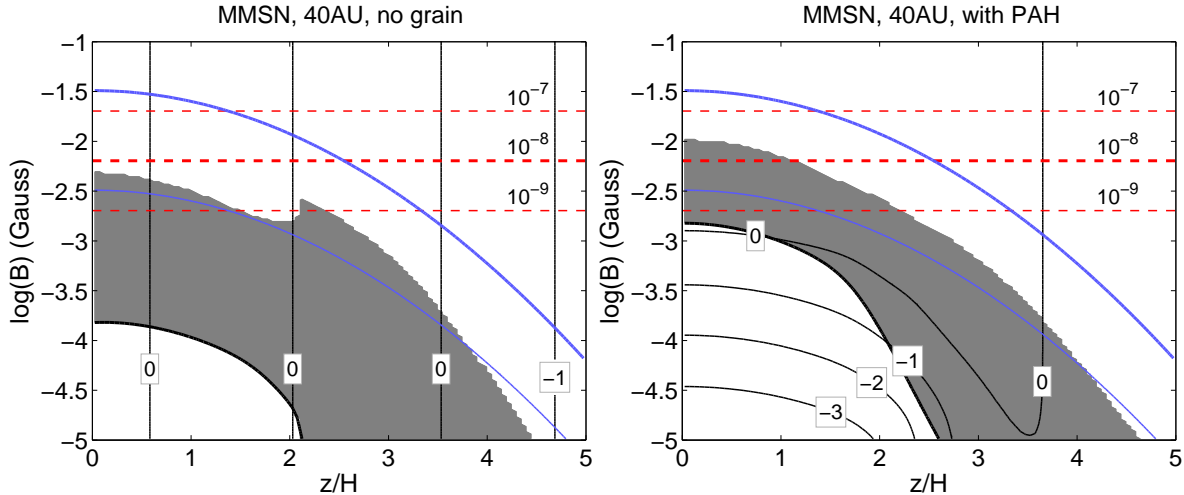}
  \caption{MRI permitted region (shaded) in the MMSN model at 40 AU from the grain-free calculation
  (left) and the calculation including PAHs with $x_{\rm PAH}=10^{-8}$ (right). Bold black solid line
  marks the boundary given by the Ohmic Elsasser number criterion $\Lambda=1$. Thin black solid
  lines show contours of constant AD Elsasser number $Am$, labeled by $\log_{10}(Am)$. Bold and
  thin blue lines mark plasma $\beta=1$ and $\beta=100$ respectively. The red dashed lines indicate
  the required field strength in the MRI permitted region in order for driving accretion rate of $10^{-7}$,
  $10^{-8}$ and $10^{-9}M_{\bigodot}$ yr$^{-1}$.}\label{fig:constrain}
\end{figure*}

Near the disk midplane, we see that $n_{\rm gr}\gg n_{\rm gr}^{\pm}\gtrsim n_i\gg n_e$. In this
regime, the electrons produced from H$_2$ ionization is quickly swallowed by the grains. Similarly,
the ions exchange charge with neutral grains to produce positively charged grains. Therefore,
ionization effectively takes place on the grains: 2 gr $\rightarrow$ gr$^+$ + gr$^-$, with the same
ionization rate $\zeta_{\rm eff}$. The dominant recombination channel is simply its inverse reaction,
with recombination rate given by equation (3) of \citet{UN90}. For recombination between two
equal sized tiny grains ($e^2/akT\gg1$), it reduces to
\begin{equation}
\begin{split}\label{eq:rcbrate}
\langle\sigma v\rangle_{gg}&\approx\bigg(\frac{48e^4}{\rho_dakT}\bigg)^{1/2}\\
&\approx2.5\times10^{-8}a_1^{-1/2}T_{100}^{-1/2} {\rm s}^{-1}{\rm cm}^{3}\ ,
\end{split}
\end{equation}
where $\rho_d=3$ g cm$^{-1}$ is the grain mass density, $k$ is the Boltzmann constant, $T$
is the temperature. In the second equation $a_1$ denotes grain size normalized to 1 nm, and
$T_{100}=T/100$ K. As long as charged grain recombination is the dominant recombination
process, the abundance of charged grains can be approximately given by
\begin{equation}
\begin{split}\label{eq:ngrpm}
\frac{n_{\rm gr}^{\pm}}{n_H}&\approx\sqrt{\frac{\zeta_{\rm eff}}{2\langle\sigma v\rangle_{gg}n_H}}\\
&\approx1.3\times10^{-10}\zeta_{\rm eff, -17}^{1/2}\ n_{H, 10}^{-1/2}\ a_1^{1/4}\ T_{100}^{1/4}\ ,
\end{split}
\end{equation}
where $\zeta_{\rm eff, -17}$ is the ionization rate normalized to $10^{-17}\ {\rm s}^{-1}$, $n_{H, 10}$
is the number density of the hydrogen atoms normalized to $10^{10}\ {\rm cm}^{-3}$. Plugging in the
numbers relevant to Figure \ref{fig:nsp} at midplane, we find
$n_{\rm gr}^{\pm}/n_H\approx1.1\times10^{-10}$. Our chemistry calculation gives $6.5\times10^{-11}$
for the averaged abundance of charged grains, which is slightly smaller due to small contributions
from other recombination channels, but is within a factor of 2 from the analytical estimate. Our
chemistry calculations further reveal that $n_{\rm gr}^{\pm}/n_H$ increases weakly with total grain
abundance $n_{\rm gr}$ (or $x_{\rm PAH}$) and approaches the asymptotic value (\ref{eq:ngrpm}).

In the grain-free case, the electron abundance $n_{e0}$ is determined by the balance between
ionization and multiple recombination channels, dominated by dissociative recombinations.
Typical electron-ion dissociative recombination rate coefficients are on the order of $10^{-7}$
s$^{-1}$ cm$^{-3}$ at $100$K, which is a factor of several higher than the grain recombination
coefficient (\ref{eq:rcbrate}). The higher recombination rate leads to smaller ionization level than
our estimate (\ref{eq:ngrpm}), which explains why $n_{e0}<\bar{n}$ in the presence of abundant
tiny grains. We note that the value of $n_{e0}$ depends on the choice of chemical reaction
network. Simple reaction network (such as \citealp{OD74}) generally produces larger $n_{e0}$
mainly because of the lack of recombination channels. In our complex (and presumably more
realistic) network, we find that the dominant recombination process is due to NH$_4^+$ and
CH$_3$CNH$^+$ in this particular case, giving $n_{e0}/n_H$ to be about $3.2\times10^{-11}$.
As a result, we obtain $\bar{n}/n_{e0}\approx4$ in the midplane, which leads to a substantial
net reduction of the AD coefficient.

\subsection[]{Active Layer and Accretion Rate}

The fact that $\bar{n}>n_{e0}$ implies that tiny grains may facilitate the MRI by suppressing AD. To
see this more explicitly, we calculate the non-ideal MHD diffusion coefficients and apply the criteria
(20) in \citet{Bai11a} to identify the MRI-active regions in our adopted PPD model. Briefly, for the
MRI to operate, the Ohmic Elsasser number $\Lambda\equiv v_A^2/\eta_O\Omega$ has to be
greater than unity, where $v_A=B/\sqrt{4\pi\rho}$ is the Alfv\'en velocity. Moreover, the magnetic
field has to be weaker than some critical value, which is set by $\beta\geq\beta_{\rm min}(Am)$.
Here plasma $\beta$ is the ratio of gas to magnetic pressure (not to be confused with the Hall
parameter), $Am\equiv v_A^2/\eta_A\Omega$ is the AD Elsasser number, and
$\beta_{\rm min}(Am)$ is given by \citep{BaiStone11a}
\begin{equation}
\beta_{\rm min}=\bigg[\bigg(\frac{50}{Am^{1.2}}\bigg)^2
+\bigg(\frac{8}{Am^{0.3}}+1\bigg)^2\bigg]^{1/2}\ ,\label{eq:betamin}
\end{equation}
which increases with decreasing $Am$. For the number density profile given in Figure \ref{fig:nsp},
the result is shown in Figure \ref{fig:constrain}.

We see that the inclusion of tiny grains strongly increases the Ohmic resistivity, as expected. At
disk midplane, magnetic field has to be about 10 times stronger than the grain-free case in order
for $\Lambda$ to be above 1 (i.e., $\eta_O$ is about 100 times larger\footnote{Note that since
$\bar{n}\approx10^3n_e$ as read from Figure \ref{fig:nsp}, electrons and other charged species
contribute roughly equally to the Ohmic conductivity.}). However, the ionization fraction at disk
midplane is still large enough such that in both cases, the required field strength for $\Lambda>1$
is still far from equipartition (i.e., $\beta\gg1$). This fact makes Ohmic resistivity essentially
irrelevant in determining the vertical extent of the active layer and the accretion rate (since the
midplane is already active), while the decisive role is played by AD.

In Figure \ref{fig:constrain}, contours of constant $Am$ are straight lines in the grain free case, as
$Am$ is independent of the magnetic field. With the inclusion of tiny grains, we see that near the
midplane region, $Am$ increases by $3$ orders of magnitude as magnetic field increases. This
corresponds to the transition region in Figure \ref{fig:model} with $\beta_i<1<\beta_e$. Eventually,
at sufficiently strong magnetic field ($\beta_i>1$), $Am$ becomes independent of magnetic field
again. Since $\bar{n}>n_{e0}$, $Am$ is larger in the presence of PAHs than in the grain-free case.
Therefore, according to (\ref{eq:betamin}), stronger magnetic field is permitted near the disk
midplane thanks to the PAHs. According to the strong linear correlation between the
Shakura-Sunyaev $\alpha$ parameter and the magnetic field strength \citep{BaiStone11a}, an
approximate magnetic field strength can be derived if MRI operates in PPDs \citep{Bai11a}
\begin{equation}
B\approx2\sqrt{\dot{M}\Omega/H}\approx1.0\dot{M}_{-8}^{1/2}r_{\rm AU}^{-11/8}\ {\rm G}\ ,
\end{equation}
which assumes the thickness of the active layer to be $H$ on each side, and the second equation
assumes one solar mass protostar with MMSN temperature profile. Here $\dot{M}_{-8}$ is accretion
rate measured in $10^{-8}M_{\bigodot}$ yr$^{-1}$, $r_{\rm AU}$ is disk radius measured in AU.
The required magnetic field strengths for typical observed accretion rates are also illustrated in the
Figure. Since stronger field leads to faster accretion, it becomes clear that PAHs are able to make
PPDs accrete more rapidly than in the grain-free case.

We note that the above explanation for the enhancement of accretion by tiny grains no longer
completely holds in the inner region of PPDs ($\lesssim10$ AU), where the disk surface density is
far above the X-ray penetration depth and accretion is layered. In such situations, the lower boundary
of the active layer is determined by both Ohmic resistivity and AD for the two Elsasser number
criteria to be satisfied (e.g., see the left panel of Figure \ref{fig:transition} in the Appendix). Since
Ohmic resistivity is larger in the presence of grains (see the end of Section \ref{sec:grain}), accretion
is thus less efficient than the grain-free case. On the other hand, when PAHs are sufficiently
abundant, the increase in Ohmic resistivity is bounded ($\lesssim10^3$) since conductivity is
dominated by the charged PAHs rather than electrons, and the reduction in accretion rate is in
general no more than the case with $0.1\mu$m grains (see the next subsection).  

\subsection[]{Parameter Study}\label{ssec:parameter}

\begin{figure*}
    \centering
    \includegraphics[width=160mm]{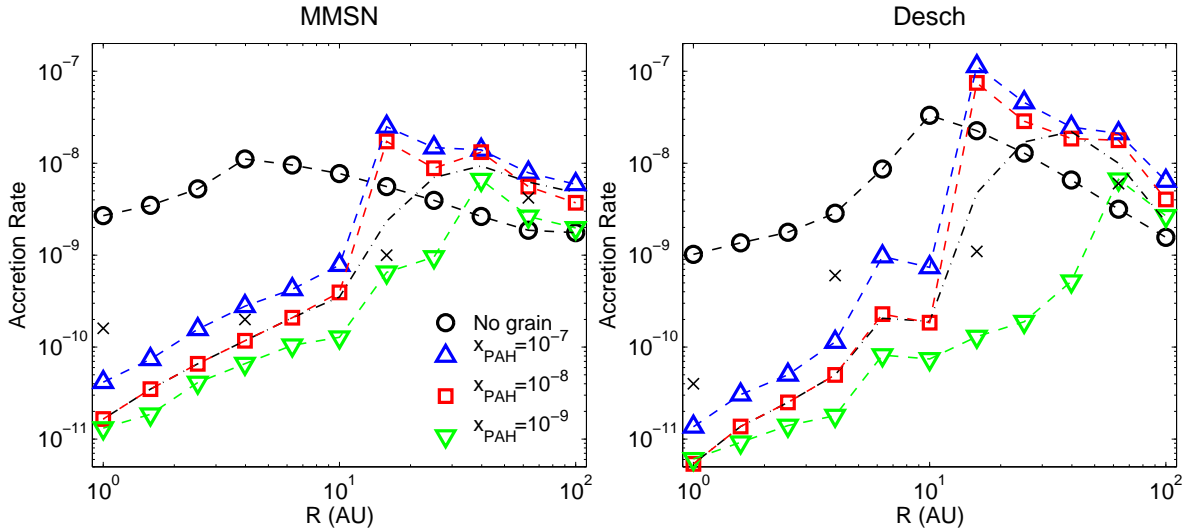}
  \caption{Optimistically predicted accretion rate as a function of disk radius for the MMAN disk (left)
  and the Desch's disk (right) models. In each panel, we show results from grain-free calculations
  (black circles) and calculations with various PAH abundances as indicated in the legend. For
  $x_{\rm PAH}=10^{-8}$, we also show the results from calculations without cosmic-ray ionization
  in the black dash-dotted line. The cross symbols in each panel denote results from the same set
  of calculations at selected radii but with $x_{\rm PAH}=10^{-11}$.}\label{fig:acrate}
\end{figure*}

A more quantitative estimate of the disk accretion rate is discussed in \citet{Bai11a} (see their
Section 4.3). The predicted accretion rate $\dot{M}$ is based on the hypothesis that MRI tends to
arrange the field configuration to maximize the rate of angular momentum transport, and is optimistic.
To demonstrate the role of tiny grains more extensively, we perform our calculations for both the
MMSN model and the more massive Desch's disk model at a wide range of disk radii from 1 AU to
100 AU, and consider PAH abundances of $x_{\rm PAH}=10^{-9}$, $10^{-8}$ (fiducial) and
$10^{-7}$. In addition, we conduct one calculation with cosmic-ray ionization turned off at
$x_{\rm PAH}=10^{-8}$ for both disk models. In Figure \ref{fig:acrate}, we show the predicted $\dot{M}$
as a function of disk radius for all our calculations. This Figure is to be compared with Figure 4 of
\citet{Bai11a}, who considered the cases with $0.1\mu$m and $1\mu$m grains. We see that the
inclusion of tiny grains makes the predicted $\dot{M}$ comparable to that with solar abundance
$0.1\mu$m grains at 1 AU, which is one to two orders magnitude smaller than the grain-free case.
However, at larger disk radii, tiny grains promote accretion compared with sub-micron grains, with
predicted rate that even exceeds the grain-free case. Below we discuss the effect of PAHs in more
detail.

We find that whenever the active layer extends to the disk midplane (the gray area touches the
vertical axis in Figure \ref{fig:constrain}, which makes the Ohmic resistivity irrelevant), the predicted
$\dot{M}$ in the presence of PAHs exceeds the grain-free case, and the reason is simply due to the
reduction of AD coefficients as discussed in the previous two subsections. This occurs at disk radii
$r\gtrsim r_{\rm trans}\approx15$ AU for $x_{\rm PAH}=10^{-7}$ and $10^{-8}$, and for
$r\gtrsim r_{\rm trans}\approx40$ AU for $x_{\rm PAH}=10^{-9}$. Inside the transition radius
$r_{\rm trans}$, accretion is layered and the predicted $\dot{M}$ is well below that in the grain-free 
case due to Ohmic resistivity as discussed at the end of the previous subsection. The increase in
predicted $\dot{M}$ near $r_{\rm trans}$ is very sharp, which is closely related to dependence of
$Am$ on magnetic field strength shown in Figure \ref{fig:model}, and is discussed in more detail in
the Appendix. 

An interesting and counterintuitive fact is that higher PAH abundance leads to faster accretion at
all disk radii. This fact can
be understood as follows. First of all, near the base of the active layer, we find that charged grain
abundance is so much higher than the electron abundance that the Ohmic resistivity is determined
by charged grains rather than electrons (i.e., $\theta\gtrsim1$). Therefore, even the abundance of
free electrons rapidly decreases with increasing $x_{\rm PAH}$, the Ohmic resistivity approximately
stays at the same level (while still much larger than the grain-free resistivity).
Secondly, as we mentioned in the discussion after equation (\ref{eq:ngrpm}), the grain ionization
level $n_{\rm gr}^{\pm}/n_H$ increases weakly with the total grain abundance $x_{\rm PAH}$. This
fact reduces both the Ohmic resistivity and the AD coefficient, which leads to faster accretion. Only
at $r\lesssim1$ AU, and when $x_{\rm PAH}$ is small ($10^{-9}$ or less), does the Ohmic resistivity
at the base of the active layer determined by electrons, which produces higher accretion rate than
larger $x_{\rm PAH}$ cases.

We see that in the presence of PAHs, the predicted $\dot{M}$ increases with disk radius for
$r\lesssim r_{\rm trans}$ where accretion is layered, and falls off with radius for
$r\gtrsim r_{\rm trans}$ as the disk midplane is activated. Also, larger disk surface density
generally leads to smaller $\dot{M}$ in the inner disk when accretion is layered (although
not by much), while when the disk midplane becomes active in the outer disk, the Desch's disk
model gives higher $\dot{M}$. These features are all consistent with the results presented in
\citet{Bai11a}, where explanations are offered (see Section 5.1).

In addition, we see in the dash-dotted line of Figure \ref{fig:acrate} that when cosmic-ray ionization
is turned off (since it may be shielded by protostellar winds), the predicted $\dot{M}$ is almost
indistinguishable with our fiducial case at the inner 10 AU. The predicted $\dot{M}$ still exceeds that
in the grain-free case in the outer disk, with the transition radius $r_{\rm trans}$ slightly larger than
that in the fiducial calculation, and the increase in $\dot{M}$ near $r_{\rm trans}$ is less sharp.
Beyond $r_{\rm trans}$, the accretion rate reduction is generally within a factor of $2$ of the fiducial
result. These results are all consistent with \citet{Bai11a}, where it is shown and discussed that
the deeper-penetrating cosmic-ray ionization does not substantially enhance the accretion rate
compared with the X-ray ionization (see Section 5.2).

Finally, we emphasize that the existence of $r_{\rm trans}$ and the fact that $\dot{M}$ increases
with $x_{\rm PAH}$ holds only when PAHs are sufficiently abundant ($x_{\rm PAH}\gtrsim10^{-9}$).
For comparison, we also show in Figure \ref{fig:acrate} with cross symbols the predicted accretion
rate at selected disk radii with $x_{\rm PAH}=10^{-11}$. At small radius of $\sim1$ AU, the predicted
$\dot{M}$ is well above the cases with higher PAH abundances. This is because the free electrons
is much more abundant so that electron conductivity dominates grain conductivity, and gives smaller
resistivity. The disk midplane becomes active at $r\gtrsim15$ AU, but we do not find the sharp jump
in the predicted $\dot{M}$ as in the case with higher PAH abundances, and the predicted accretion
rate is also smaller by a factor of up to 100. This is because $x_{\rm PAH}$ is so small that the
grain number density is well below $n_e$ (so $\bar{n}/n_e\approx1$) and the effects we discussed
before no longer apply.

The role of PAHs in PPDs has been recently studied by \citet{PerezBeckerChiang11}.
Besides the fact that they used a more stringent criterion for AD to suppress the MRI, the $Am$
values they obtained were largely underestimated because they adopted the simple formulae
(\ref{eq:diff0}) without taking into account the contribution from grains. This has led them to
conclude that the presence of PAHs makes the MRI too inefficient to drive rapid accretion at all
disk radii, although as we have shown that PAHs in fact promote accretion in the outer disk, and
at all disk radii the predicted accretion rate generally increases with PAH abundance. Our findings
are subsequently confirmed in their follow-up work \citep{PerezBeckerChiang11b} with corrected
grain conductivities.

\subsection[]{Effect of Grain Size Distribution}

\begin{figure}
    \centering
    \includegraphics[width=90mm]{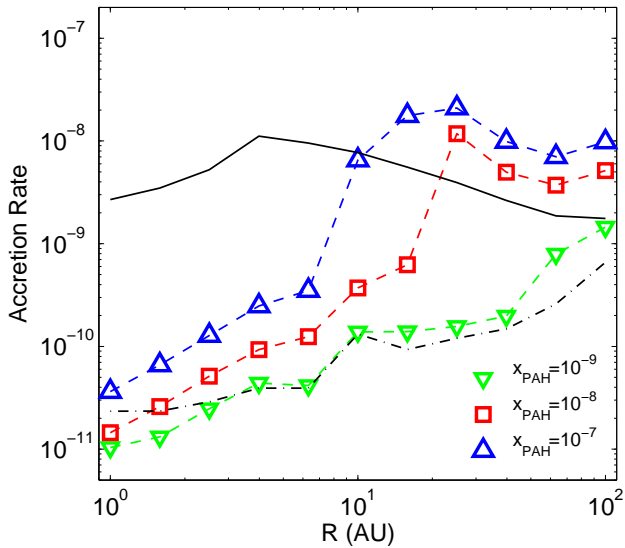}
  \caption{Optimistically predicted accretion rate as a function of disk radius for the MMSN disk with
  two populations of grains: $0.1\mu$m grains with fixed mass fraction of $1\%$ and PAHs with
  abundance $x_{\rm PAH}=10^{-9}, 10^{-8}$ and $10^{-7}$. Comparison is made to the grain-free
  case (black solid) and case with single population of $0.1\mu$m grains (black dash-dotted).}
  \label{fig:acmix}
\end{figure}

Our study so far have considered only single-species of grains. In reality, tiny grains coexist with
larger grains with the latter dominates the grain mass (but not necessarily abundance). To justify
our findings in a more realistic context, we conduct another set of calculations with two
populations of grains. Besides the PAH population with $x_{\rm PAH}=10^{-9}, 10^{-8}$ and
$10^{-7}$ that contributes a negligible amount of mass, we add the second population of
$0.1\mu$m grains with fixed mass fraction of $1\%$. In Figure \ref{fig:acmix}, we show the predicted
accretion rate as a function of disk radius in the same way as Figure \ref{fig:acrate}. For comparison,
we also plot the predicted $\dot{M}$ in the grain-free case (black solid) and the pure $0.1\mu$m grain
case (black dash-dotted) taken from \citet{Bai11a}. We see that at fixed PAH abundance with
$x_{\rm PAH}\gtrsim10^{-8}$, the predicted $\dot{M}$ with two grain populations is very close to that
in our previous calculations with a single PAH population, and it behaves as if the $0.1\mu$m grains
do not exist. In particular, the sharp enhancement of accretion beyond the transition radius is clearly
present, and the predicted $\dot{M}$ increases with $x_{\rm PAH}$. At $x_{\rm PAH}=10^{-9}$, the
situation is marginal: with the addition of $0.1\mu$m grains, the signature of sharp transition is less
obvious, which occurs at $r_{\rm trans}\sim60$ AU, and beyond $r_{\rm trans}$ the predicted
$\dot{M}$ is still less than that in the grain-free case. In fact, the behavior of the two-population
calculation at $x_{\rm PAH}=10^{-9}$ roughly lies in between the two single-population limits.

Our results are consistent with chemistry studies by \citet{BaiGoodman09} which indicate that
different grain populations tend to behave independently. Therefore, the outcome of a size
distribution of grains is mainly determined by the grain population that dominates the
recombination process. \citet{BaiGoodman09} found that the controlling parameter of the
recombination lies in between the total surface area and the grain abundance weighted by
linear size. As an approximation we weigh the grain abundance by $a^{3/2}$ and find that PAHs
and $0.1\mu$m sized grains contribute roughly equally to the recombination process when
$x_{\rm PAH}\sim3\times10^{-9}$, consistent with our findings discussed in the previous
paragraph.

In sum, the effect of tiny grains studied in this paper is approximately independent of the presence
of larger grains, and we quote $x_{\rm PAH}=10^{-9}$ as the critical tiny grain abundance above
which enhancement of accretion is possible in the outer regions of PPDs.

\section[]{Summary and Discussion}\label{sec:conclusion}

Tiny grains are important charge carriers in weakly ionized gas and strongly affect the non-ideal
MHD effects when their abundance is above the ionization fraction. We generalize the commonly
used simple expressions for non-ideal MHD diffusion coefficients (\ref{eq:diff0}) to incorporate the
effect of charged tiny grains. We show that at sufficiently small ionization level, tiny grains become
the dominant charge carriers, and the non-ideal MHD diffusion coefficients behave very differently
from the grain-free case. The Ohmic conductivity is dominated by charged grains rather than
electrons when $\bar{n}$ exceeds about $10^3n_e$. In the AD regime (strong magnetic field), Hall
and AD coefficients are strongly reduced by a factor of about $(\bar{n}/n_e)^2$ relative to those in
the Ohmic regime (weak magnetic field), and for sufficiently large $\bar{n}/n_e$, Hall dominated
regime diminishes.

We study the role of tiny grains on the MRI driven accretion in PPDs, and find that novel behaviors
occur when the tiny grains are sufficiently abundant with $x_{\rm PAH}\gtrsim10^{-9}$, regardless
of whether larger grains are present or not. At the inner
disk where accretion is layered, the predicted accretion rate in the presence of tiny grains is one to
two orders of magnitude less than the grain-free case due to increased Ohmic resistivity, but is
similar to or higher than that with solar-abundance $0.1\mu$m grains. A sharp increase in the
predicted $\dot{M}$ occurs at the transition radius $r_{\rm trans}\approx15$ AU (in the fiducial
model) where the disk midplane becomes active, making Ohmic resistivity irrelevant to the accretion
rate. Quite unexpectedly, we find that at $r\gtrsim r_{\rm trans}$, tiny grains make
accretion even more rapid than the grain-free case. Moreover,
our predicted accretion rate increases with PAH abundance. These results are due to that at
large PAH abundance, ionization-recombination balance makes $\bar{n}$ orders of magnitude
larger than $n_e$, and even exceeds the grain-free electron density $n_{e0}$ at disk midplane.
These facts prevent Ohmic resistivity from rapidly increasing as $x_{\rm PAH}$ increases, reduce
the dissipation by AD, thus facilitate the active layer to extend deeper into the disk midplane and
permit stronger MRI turbulence. Our results highlight the importance of evaluating the full
conductivity tensor in the calculation of the non-ideal MHD diffusion coefficients rather than using
the simple grain-free formulae (\ref{eq:diff0}).

We emphasize that the effects studied in this paper mainly apply to tiny grains
($\lesssim0.01\mu$m), and when they are sufficiently abundant ($x_{\rm PAH}\gtrsim10^{-9}$). For
grains larger than $0.1\mu$m, their abundance is at most $3\times10^{-12}$ per H$_2$ molecule
(for having 1$\%$ of mass), which is orders of magnitude smaller than $10^{-9}$. The reduction of
AD coefficient also exist for these relatively larger grains, as can be seen in the bottom panels of
Figure 3 in \citet{Bai11a}, but its effect is much more limited than the tiny grain case. When large
grains and tiny grains coexist, the effect of large grains becomes negligible if $x_{\rm PAH}>10^{-9}$.

Although tiny grains strongly enhance PPD accretion in the outer disk, the situation at the inner disk
is still similar to the case with $0.1\mu$m grains \citep{Bai11a}, with predicted accretion rate much
less than the typical value of $10^{-8}M_{\odot}$ yr$^{-1}$ as inferred from observations
\citep{Hartmann_etal98}. The recently proposed far ultraviolet (FUV) ionization scenario does not
provide large accretion rate in the inner disk either due to the small penetration depth of FUV photons
\citep{PerezBeckerChiang11b}. Therefore, either additional ionization sources has to make the MRI
more efficient than the X-rays and FUV photons (candidate may include energetic protons from the
protostars \citep{TurnerDrake09}), or additional mechanism such as magnetized wind
\citep{Salmeron_etal07} operates to provide the angular momentum transport in the inner disk.

In the case of transitional disks characterized by inner holes or gaps
\citep{Calvet_etal02,Espaillat_etal07}, we note that the observationally inferred outer boundary of
the holes or gaps is typically at a few tens of AU \citep{Hughes_etal09, Kim_etal09}, which is
generally greater than $r_{\rm trans}$ in our models with PAHs. Therefore, the enhancement of
accretion by tiny grains works effectively for transitional disks, and X-ray driven MRI with PAHs
is able to feed the inner hole /gap at the (optimistic) rate of about $10^{-8}M_{\odot}$ yr$^{-1}$.
This is sufficient to account for the observed accretion rate in transitional disks
\citep{Najita_etal07,Sicilia_etal10}, with the accreting gas fed from the outer disk flowing through
the hole / gap possibly guided by multiple planets \citep{CMC07,PerezBeckerChiang11,Zhu_etal11}.

\acknowledgments

X.-N.B thanks Jim Stone and Jeremy Goodman for carefully reading the manuscript with
comments, and Bruce Draine for useful discussions on PAHs. Comments from Daniel Perez-Becker
and the referee, Eugene Chiang, are especially acknowledged which lead to several improvements to
this work. This work is supported by NASA Headquaters under the NASA Earth and Space Science
Fellowship Program Grant NNX09AQ90H awarded to X.N.B.

\appendix

\section[]{Midplane Activation at Transition Radius}

\begin{figure*}
    \centering
    \includegraphics[width=160mm]{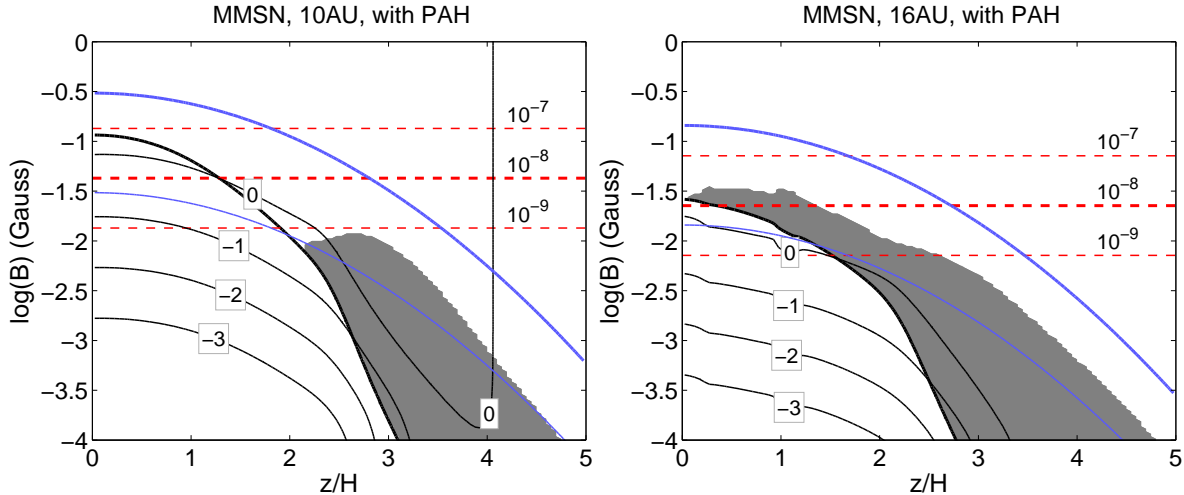}
  \caption{Same as Figure \ref{fig:constrain}, but for MRI permitted region (shaded) in the
  MMSN model at 10 AU (left) and 16 AU (right) from calculations with PAHs at
  $x_{\rm PAH}=10^{-8}$.}\label{fig:transition}
\end{figure*}

It has been shown in Figure \ref{fig:acrate} and discussed in Section \ref{ssec:parameter} that
near the transition radius $r_{\rm trans}$ where the disk midplane becomes active, the predicted
accretion rate $\dot{M}$ increases sharply with radius, from well below the grain-free rate at
$r<r_{\rm trans}$ to a factor of several higher at $r>r_{\rm trans}$. We discuss the activation of the
midplane by the MRI in this Appendix and explain the sharp dependence of $\dot{M}$ on disk radius
near $r_{\rm trans}$.

Figure \ref{fig:transition} shows the MRI permitted region in our fiducial model at 10 and 16 AU
between which lies the transition radius.
We see from this plot that much stronger magnetic field is permitted in the disk midplane once it is
activated, which is the cause of the big jump in $\dot{M}$.

Looking more closely into the sharp transition, we find it is related to the dependence of the AD
Elsasser number $Am$ on the magnetic field strength. We have shown in Figure \ref{fig:model}
that $\eta_A$ depends on the magnetic field quadratically in weak ($\beta_e<1$) and strong
($\beta_i>1$) field regimes as commonly considered, but at the intermediate regime with
$\beta_i<1<\beta_e$, $\eta_A$ behaves similarly as $\eta_O$ and does not depend on the magnetic
field strength. Consequently, the contours of Ohmic and AD Elsasser numbers $\Lambda$ and
$Am$ are nearly parallel to each other, and both shapes are close to (near the midplane) or slightly
steeper (near the surface) than the contours of constant $\beta$, as can be seen in Figure
\ref{fig:transition}. In the mean time, an increase of ionization level (as one moves to outer radius)
brings down the $Am$ and $\Lambda$ contours relative to the $\beta$ contour in a wide range of
disk heights. Then, according to our criteria for the MRI to operate ($\Lambda>1$ and
$\beta>\beta_{\rm min}(Am)$), once the ionization level in the disk reaches some critical degree
(i.e., left panel of Figure \ref{fig:transition}), a further increase in ionization would suddenly make both
criteria satisfied simultaneously across the midplane (i.e., right panel of FIgure \ref{fig:transition}),
leading to the sharp transition.

\bibliographystyle{apj}

\label{lastpage}
\end{document}